\documentclass[preprint,5p,twocolumn]{m-elsarticle}
\usepackage{amsmath}
\usepackage{amssymb}
\usepackage{epsfig}
\usepackage{color}
\definecolor{bl}{cmyk}{1,1,0.3,0}
\definecolor{gr}{rgb}{0,0.35,0.1}
\definecolor{new}{rgb}{0.21,0.43,0.5}

\usepackage{hyperref}
\hypersetup{colorlinks=true,urlcolor=new,linkcolor=new,citecolor=new}
\newcommand{\publish}[1]{\textit{To appear in:}\,#1}
\begin{document}

%%%%%%%%%%%%%%%%%%%%%%%%%%%%%%%%%%%%%%%%%%%%%%%%%%
\begin{frontmatter}
%%%%%%%%%%%%%%%%%%%%%%%%%%%%%%%%%%%%%%%%%%%%%%%%%%
\title{\textsf{Anti-gravity and/or dark matter contributions from massive gravity}}
%%%
\author[ulb]{Michael V. Bebronne}
\address[ulb]{Service de Physique Th\'eorique, Universit\'e Libre de Bruxelles (ULB),\\ CP225, boulevard du Triomphe, B-1050 Bruxelles, Belgium.}
%%%
\begin{abstract}
Recently, the static spherically symmetric solution of the gravitational field equations have been found in theories describing massive graviton with spontaneous breaking of the Lorentz invariance \cite{Bebronne:2009mz}. These solutions, which show off two integration constants instead of one in General Relativity, are discussed. They are candidates for modified black holes provided they are stable against small perturbations. These solutions may have both attractive or repulsive behavior at large distances. Therefore, these modified black holes may mimics the presence of dark matter or be a source of anti-gravity.
\end{abstract}

%%%%%%%%%%%%%%%%%%%%%%%%%%%%%%%%%%%%%%%%%%%%%%%%%%
\end{frontmatter}
%%%%%%%%%%%%%%%%%%%%%%%%%%%%%%%%%%%%%%%%%%%%%%%%%%

\section{Introduction}

General Relativity describes the gravitational interaction through the exchange of massless particles, the gravitons. But could Einstein's theory be generalized as to describe massive graviton? Since the original work of \emph{W. Pauli} and \emph{M. Fierz} \cite{Pauli:1939xp} in 1939, the attempts to answer this question have spark off lots of attention, and still the debate is far from being closed (for a review on the subject, see \cite{Rubakov:2008nh}). Apart from the theoretical challenge in modifying Einstein's theory, recent advances in observational cosmology \cite{Dunkley:2008ie} has revived interest in large scale modifications of General Relativity. Theories of massive gravity, describing massive gravitons, belong to this category.

Interestingly, on galactic and cosmological scales, the predictions of General Relativity theory actually do not agree with observations; the agreement is only achieved after the introduction of the otherwise undetected dark matter and dark energy. Yet, another explanation to these dark paradigms may be possible: the law governing gravity on large scale could be different from expected \footnote{A combination of both explanations might also be needed.}. Hence, in parallel with the direct searches for the dark components, alternatives to General Relativity should be explored.

From a phenomenological point of view, it is legitimate to study massive gravity since the constraints on an hypothetical graviton mass are much weaker than those on massive \mbox{photon \cite{Goldhaber:2008xy}}. Indeed, up to now gravitational waves have not been directly observed, although the secular decrease of orbital period of binary pulsars has been shown to be compatible with the emission of gravitational wave as predicted by General Relativity \cite{1979Natur.277..437T}. The constraints on the graviton mass are even lower for theories \cite{Dubovsky:2005dw} in which the Lorentz invariance is spontaneously broken. These models are motivated by the consistency problems that arise when trying to define a Lorentz invariant theory of massive gravity \cite{Arkani-Hamed:2002sp}. In the context of Lorentz breaking massive gravity, the coherence between General Relativity and the observations requires the graviton mass to be smaller than the inverse period of orbital motion of binary pulsars \cite{Dubovsky:2004ud}
\begin{eqnarray*}
m \lesssim 10^{-19} \textrm{eV} \sim  \left( 10^{15} \textrm{cm} \right)^{-1} .
\end{eqnarray*}
Indeed, because of the absence of Lorentz invariance, Newton's potential remains unmodified in the linear approximation despite the non-vanishing graviton mass. The Solar System constraints are therefore satisfied for rather large graviton mass.

\section{The model}

The model under consideration is given by the following action \cite{Dubovsky:2004sg}
\begin{eqnarray*}
\mathcal{S} = \int \textrm{d}^4 x \sqrt{- g} \left[ - \textrm{M}^2_{\textrm{\footnotesize pl}} \mathcal{R} + \mathcal{L}_{\textrm{\footnotesize m}} + \Lambda^4 \mathcal{F} \left( X , W^{ij} \right) \right] .
\end{eqnarray*}
The first two terms comprise the usual General Relativity action; they are the curvature and the Lagrangian of the minimally coupled ordinary matter. The third term is a new contribution describing four scalar fields $\phi^0$ and $\phi^i$ (with $i= 1,2,3$), through the following variables
\begin{eqnarray*}
X &=& \frac{g^{\mu\nu} \partial_\mu \phi^0 \partial_\nu \phi^0}{\Lambda^{4}} \\
%%%
W^{ij} &=& \left( g^{\mu\nu} - \frac{\partial^\mu \phi^0 \partial^\nu \phi^0}{\Lambda^{4} X} \right) \frac{\partial_\mu \phi^i \partial_\nu \phi^j}{\Lambda^{4}} ,
\end{eqnarray*}
where $\Lambda$ is a UV cutoff. The four scalar fields are known as Goldstone fields since their space-time dependent vacuum expectation values break spontaneously the Lorentz invariance of the model. It has the form
\begin{eqnarray*}
g_{\mu\nu} = \eta_{\mu\nu} , & \phi^0 = \Lambda^2 t , & \phi^0 = \Lambda^2 x^i ,
\end{eqnarray*}
while keeping the invariance under three dimensional rotations intact. It is worth noting that the way these four scalar fields break the Lorentz invariance is not fundamentally different from the way the CMB actually breaks it. Indeed, any observer could determine his motion with respect to the CMB by studying the CMB dipole, and there is only one reference frame which is at rest with respect to the CMB.

Models of massive gravity described by the previous action are free of the usual pathologies that plague Lorentz invariant massive gravity. Despite the fact that they have massive gravitons, these theories have a very interesting phenomenology. For example, the prediction of these models concerning the growth of perturbations in the post-inflationary Universe are compatible with General Relativity's predictions \cite{Bebronne:2007qh}. Moreover, the density perturbations could even grow faster in these models than in the Einstein's theory. Another interesting feature is the presence of an instantaneous interaction \cite{Bebronne:2008tr}. Finally, contrary to what General Relativity predicts, black holes seems to posses hairs in Lorentz breaking massive gravity \cite{Dubovsky:2007zi}.

\section{Exact spherically symmetric solutions}

The exact spherically symmetric solution, or Schwarzschild solution, plays a central role in General Relativity. First, it describe the metric outside of spherical non-rotating bodies. In its weak field limit, this solution reduce to Newton's potential. Second, this solution describe black holes. Interestingly, the Schwarzschild solution is modified in Lorentz-violating massive gravity \cite{Bebronne:2009mz}. Indeed, consider a toy model characterized by the following function
\begin{eqnarray*}
\mathcal{F} &=& c_{0} \left[ \frac{1}{X} + w_1 \right. \\
& & \left. - \frac{\lambda}{12} \left( w_1^3 - 3 w_1 w_2 - 6 w_1 + 2 w_3 - 12 \right) \right] ,
\end{eqnarray*}
where $w_i = \textrm{Tr} \left( W^i \right)$ while $c_{0}$ and $\lambda$ are dimensionless constants. This function has been chosen in such a way that the resulting equations are solvable analytically\footnote{Another example is discussed in \cite{Bebronne:2009mz} by solving numerically the equations of motions.}. Then, the metric part of the solution reads\footnote{For this solution, the scalar fields are given by
\begin{eqnarray*}
\phi^0 &=& \Lambda^2 t \pm \Lambda^2 \int \frac{\textrm{d} r}{g_{00}} \left[ 1 - g_{00} \left( \frac{S}{c_{0} m^2} \frac{\lambda - 1}{r^{\lambda+2}} + 1 \right) \right]^{1/2}, \\
\phi^i &=& \Lambda^2 x^i .
\end{eqnarray*}}
\begin{eqnarray*}
\textrm{d} s^2 &=& \left( 1 + 2 \Phi \right) \textrm{d} t^2 - \frac{\textrm{d} r^2}{1 + 2 \Phi} - r^2 \textrm{d} \Omega^2 , \\
\Phi &=& - \left( \frac{\textrm{M G}}{r} + \frac{\textrm{S}}{2 r^\lambda} \right) .
\end{eqnarray*}
This solution depend on two integration constants, M and S, instead of one in Einstein's theory. For S = 0 this solution reduce to the usual Schwarzschild solution describing a black hole of mass M. Therefore, the ``scalar charge'' S is responsible for a modification of the geometry as compared to General Relativity. The behavior of this solution is determined by the value of the two integration constants and by the value of $\lambda$. Taking $\lambda > 1$ will guarantee that the new term, proportional to S, dominates at small distances while the usual Schwarzschild term dominates at infinity. Then, the mass measured by an observer who believes in General Relativity will converge to M at infinity.

Both terms in the potential are singular at the origin. Therefore, the physical solutions are those with an horizon to hid the singularity. The presence of an horizon depends on the relative value of M and S. If $|S|\equiv s^{-\lambda}$, the existence of an horizon requires that
\begin{eqnarray} \label{MB:eq:horizon}
s M \geq \frac{\lambda}{2G_N}\left(\frac{1}{\lambda-1} \right)^{\frac{\lambda-1}{\lambda}} .
\end{eqnarray}
The two charges M and S are determined by matching the exterior solution to the interior one, which depends of the object under consideration. For a static star made of usual matter, it is possible to show that M is actually the mass of the star while S = 0. It remains an open question how objects (e.g., black holes) with S $\neq$ 0 can be created. But it is conceivable that a non-zero scalar charge may be acquired during the gravitational collapse.

\begin{figure*}
\begin{center}
\includegraphics[width=11cm]{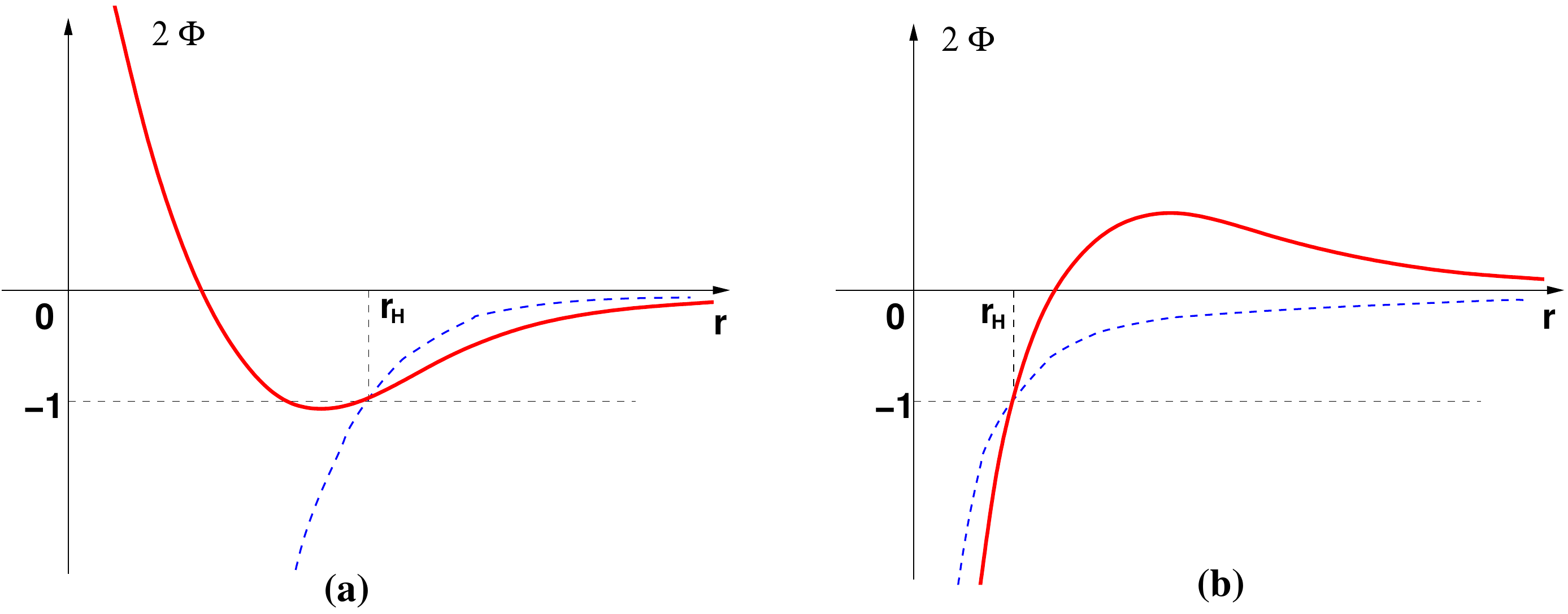}
\caption{Newton's potential for two different choices of the integration constants. Fig. (a) correspond to M $>$ 0 and S $<$ 0 while Fig. (b) represents solution with M $<$ 0 and S $>$ 0. The dashed curves are the usual Newtonian potential corresponding to the Schwarzschild solutions of General Relativity.}\label{MB:fg:potential}
\end{center}
\end{figure*}

Still, there are two interesting different types of solutions depending on the value of these charges. The first type is characterized by \mbox{M $>$ 0} and \mbox{S $<$ 0}. These solutions, shown in \mbox{Fig.\ref{MB:fg:potential}-(a)}, are attractive all the way to the horizon. Since the gravitational potential deduced from them increases slower than in the Einstein's theory, they induce a gravitational force which decreases slower than in General Relativity. Therefore, these solutions may mimic the presence of dark matter.

The second type of solutions, shown in \mbox{Fig.\ref{MB:fg:potential}-(b)}, are those which show an anti-gravitating behavior. They are characterized by \mbox{M $<$ 0} and \mbox{S $>$ 0}. These solutions are attractive below a certain distance, and become repelling at larger distances since their potential is decreasing then. It is worth noting that in General Relativity only positive value of M make sense. Indeed, for $M < 0$ the Schwarzschild solution posses a naked singularity at the origin which is physically unacceptable. Moreover, the conventional matter satisfies the null energy condition which ensures that any compact spherically-symmetric matter distribution has a positive mass. None of these arguments goes through in the case of massive
gravity. Indeed, a singularity with a negative mass could still be hidden by an horizon provided that the condition (\ref{MB:eq:horizon}) is satisfied. The positivity of energy is also not expected in massive gravity, since the vacuum breaks the invariance under time translations. In massive gravity, only the combination of the time translations with the shifts of $\phi^0$ by a constant remains unbroken.

These solutions show a richer phenomenology than the one predicted by General Relativity. Still, there are several open question about them. For example, one has to check that these solutions are stable against small perturbations, otherwise the black hole interpretation will not be possible. Another open question concern the possibility of having a non zero scalar charge. These questions, among others, deserve further studies.

\section*{Acknowledgments}

The work presented here was done in collaboration with P. Tinyakov, and has been supported by the Belgian \emph{Fond pour la formation \`a la Recherche dans l'Industrie et dans l'Agriculture (FRIA).}

\end{document}